\documentclass[journal]{IEEEtran}
\usepackage{algorithm}
\usepackage[noend]{algpseudocode}
\makeatletter
\def\BState{\State\hskip-\ALG@thistlm}
\makeatother
\usepackage{soul}
\usepackage{amsmath}
\usepackage{amssymb}
\usepackage{graphics}
\usepackage{graphicx}
\usepackage{oldlfont}
 \usepackage{newlfont}
 \usepackage{amsfonts}
 \usepackage{mathtools}
\usepackage{color}
 \usepackage{ulem}
 \usepackage{multirow}

 \newcommand{\R}{{\mathbb R}}
   \usepackage{amsthm}
 \theoremstyle{definition}

\usepackage[table]{xcolor}
\usepackage{mathrsfs}
\DeclareMathOperator{\sign}{sign}

\newcommand{\A}{\mathbf A}
\newcommand{\f}{\mathbf f}
\newcommand{\m}{\mathbf m}
\newcommand{\y}{\mathbf y}

\newcommand{\SH}{\mathbf S}

\newcommand{\V}{\mathbf v}
\begin{document}

\title{An automatic regularization method: An application for 3D X-ray micro-CT reconstruction using sparse data}

\author{Zenith Purisha$^{*}$, Sakari S. Karhula, Juuso H. Ketola, Juho Rimpel{\"a}inen, \\
Miika T. Nieminen, Simo Saarakkala, Heikki Kr{\"o}ger and Samuli Siltanen

\thanks{$^{*}$Zenith Purisha, Juho Rimpel{\"a}inen and Samuli Siltanen are with the Department
of Mathematics and Statistics, University of Helsinki, Finland. An asterisk indicates the corresponding author. E-mail: zenith.purisha@helsinki.fi}.
\thanks{Zenith Purisha is with the Department of Mathematics, Universitas Gadjah Mada, Indonesia}.
\thanks{Sakari S. Karhula is with Infotech Oulu, University of Oulu, Oulu, Finland}.
\thanks{Sakari S. Karhula, Juuso H. Ketola, Miika T. Nieminen and Simo Saarakkala are with the Research Unit of Medical Imaging, Physics and Technology, University of Oulu, Oulu Finland}.
\thanks{Heikki Kr{\"o}ger is with the University of Eastern Finland, Kuopio, Finland}
\thanks{Miika T. Nieminen and Simo Saarakkala are with the Medical Research Center, University of Oulu and Oulu University Hospital, Oulu, Finland}.
\thanks{Miika T. Nieminen and Simo Saarakkala are with the Department of Diagnostic Radiology, Oulu University Hospital, Oulu, Finland}.
}
\vspace{4mm}

\maketitle
\vspace{17mm}

\begin{abstract}
X-ray tomography is a reliable tool for determining the inner structure of 3D object with penetrating X-rays. However, traditional reconstruction methods such as FDK require dense angular sampling in the data acquisition phase leading to long measurement times, especially in X-ray micro-tomography to obtain high resolution scans. Acquiring less data using greater angular steps is an obvious way for speeding up the process and avoiding the need to save huge data sets available memory. However, computing 3D reconstruction from such a sparsely sampled dataset is very sensitive to measurement noise and modelling errors. An automatic regularization method is proposed for robust reconstruction, based on enforcing sparsity in the three-dimensional shearlet transform domain. The inputs of the algorithm are the projection data and {\it a priori} known expected degree of sparsity, denoted $0<{\mathcal C}_{pr}\leq 1$. The number ${\mathcal C}_{pr}$ can be calibrated from a few dense-angle reconstructions and fixed. Human subchondral bone samples were tested and morphometric parameters of the bone reconstructions were then analyzed using standard metrics. The proposed method is shown to outperform the baseline algorithm (FDK) in the case of sparsely collected data. The number of X-ray projections can be reduced up to 10\% of the total amount while retaining the quality of the reconstruction images and of the morphometric paramaters.
\end{abstract}

\section{Introduction}

X-ray micro-tomography ($\mu$CT) is an important tool in medical imaging and in industrial computed tomography. The principle of X-ray $\mu$CT is to reveal the inner structure of an unknown object without destroying it by propagating X-rays through the object. X-ray $\mu$CT will image an internal three-dimensional (3D) structure of the object at a high resolution.

%Bone morphometric studies using X-ray micro-tomography ($\mu$CT) have recently gained importance in biomedical research. 

In $\mu$CT a set of projection images  is collected from many directions. A mathematical reconstruction algorithm is used for revealing the 3D structure inside the sample. Conventional reconstruction methods such as Feldkamp-Davis-Kress type (FDK) \cite{feldkamp1984practical} require densely sampled datasets to achieve sufficient reconstruction quality. More precisely, denote the projection angles by $\theta, 2\theta, \dots, N\theta=180,$ with a fixed angular step $\theta>0$. The FDK algorithm typically needs at least 300 projections ($N=300$) in order to deliver a high enough reconstruction quality. This often leads to the impractically of saving big data sets and long measurement times. In practice, quite often only a few X-ray $\mu$CT machines are available while the demand to get the acquisition data is high.    

%Moreover, in longitudinal  {\it in vivo} animal studies the cumulative radiation dose can modify the properties of  living tissue, thereby biasing the results \cite{boone2004small}. 

A simple way to speed up the acquisition process is to collect less data by decreasing $N$ and enlarging the angular step $\theta$ accordingly so that the 180 degree half-circle is still sampled. However, the task of computing a 3D reconstruction from such a sparsely sampled dataset becomes extremely sensitive to modelling errors and measurement noise. In mathematical terms, it is an {\it ill-posed inverse problem} \cite{mueller2012linear} that needs to be regularized by making use of {\it a priori} information about the sample structure. 

The {\it shearlet transform} is a tool for orien\-tation-aware multiscale signal processing \cite{kutyniok2012,kutyniok2016shearlab}. Shearlets  provide efficient representations for a variety of  signals, see for example \cite{Guo2012}. We implement a 3D tomographic reconstruction algorithm regularized by promoting the sparsity of the bone structure in the shearlet transform domain. We use an iterative soft thresholding algorithm, the so-called Primal Dual Fixed Point (PDFP) as outlined in \cite{chen2016primal}. The method was one of the earliest introduced in \cite{Daubechies2004}. Implementation of this method using wavelet transform as the penalty  term has been successfully studied in \cite{purisha2017}. 

We introduce a novel technique to make the regularized reconstruction process fully automatic. Namely, the PDFP method involves a thresholding parameter $\mu>0$ . All shearlet coefficients smaller than $\frac{\mu}{2}$ in absolute value are set to zero in each iteration. How can a suitable value for $\mu$ be chosen? 

If $\mu$ is large, then many coefficients vanish and the reconstruction is very sparse in the shearlet domain. If $\mu$ is small, then almost all shearlet coefficients of the final reconstruction will be nonzero. We propose determining the typical ratio $0<{\mathcal C}_{pr}\leq 1$ of nonzero shearlet coefficients from a few dense-angle 3D reconstructions of both healthy and osteoarthritic (OA) bone samples. We let $\mu=\mu_j$ change in each iteration and apply a simple control algorithm so that $\mu_j$ converges to a limit value producing a reconstruction having the {\it a priori} known sparsity ${\mathcal C}_{pr}$. The integral controller, part of proportional-integral-derivative controller (PID controller), is implemented in this approach \cite{aastrom1995pid}. 

In this work, we investigate the reliability of a modern sparsity-promoting 3D reconstruction algorithm to reconstruct human trabecular bone (healthy and OA) using sparse X-ray tomographic data. We quantify morphometric parameters of human trabecular bone calculated from 3D reconstructions ({\it e.g} the percentage of bone volume ({\it BV/TV}), trabecular thickness ({\it Tb.Th}), trabecular separation ({\it Tb.Sp}))\cite{boutroy2005vivo,feldkamp1989direct,kuhn1990evaluation,Rüegsegger1996}.

These parameters are used for validation. %important descriptors of trabeculae tissue health. 
They are important parameters to see the changes in the 3D structure of bone caused by osteoarthritis, such as subchondral bone sclerosis  \cite{bouxsein2010guidelines, postnov2003quantitative,finnila2016association,mohan2011application}. We use Computed Tomography Analyzer (CTAn) software to calculate the morphometric parameters from the trabecular bone. The parameters are defined for binary (bone/not-bone) 3D reconstructions; we use the Otsu algorithm in CTAn for segmentation  \cite{otsu1975threshold}.

Reconstructions using 3D shearlet-sparsity regularization have been shown to outperform FDK in the case of sparsely collected data. With the shearlet-sparsity reconstruction method, the number of X-ray projections can be reduced to 10\% of the currently used amount while retaining the quality of morphometric analysis. 

Shearlet-based methods for X-ray tomography have been studied before, starting with \cite{colonna2010radon} concentrating on inversion from noisy, densely sampled 2D sinograms.  Total variation regularization and shearlet sparsity have been successfully combined for 2D tomographic data in \cite{vandeghinste2012combined, garduno2016computerized}, including sparse data with a minimum of 128 angles. Shearlets have been shown to be useful for 2D region-of-interest tomography in \cite{bubba2016roi} and for limited-angle tomography in \cite{Frikel2013}. %For 3D X-ray tomography case, 3D-shearlet has been implemented in \cite{Bubba2017}.

A study to recover bone structure in 2D from sparse microtomography data using a tomographic method called the discrete algebraic reconstruction technique (DART) was introduced in \cite{batenburg2006discrete,batenburg2011dart}. The study of in vivo small animal bone in 3D was done as well in \cite{van2017discrete}. However, the method proposed here is based on different assumptions about {\it a priori} knowledge. DART needs rather accurate attenuation value estimates as input and assumes that the target consists of a small number of possible materials.

Regarding 3D tomography, the shearlet study showing the optimality for representing tomographic data in terms of shearlets is done in \cite{Guo2013}. Research in shearlet-based regularization in sparse dynamic tomography has been studied in \cite{Bubba2017}. However, the present work promotes an {\it a priori} known level of sparsity and an adaptive method for choosing regularization parameter. %To the best our knowledge, the present work is the first application of the 3D shearlet transform for measured X-ray tomographic data.

\section{Materials and Methods}\label{Materials and Methods}

\subsection{3D Tomographic Setup}\label{3D Tomographic Setup}

The goal of X-ray tomography is to recover the density function of an unknown object from measured projection data. In this paper, the object is three-dimensional, and cone-beam geometry is used for modelling the measurement.

Consider a physical domain $\Omega \subset \R^3$ and a non-negative X-ray attenuation function $f:\Omega\rightarrow\R$. The X-rays travel through $\Omega$ along straight lines $L \subset \Omega$. After calibration, each pixel value in the collection of digital radiographs yields a line integral $\int_L f(x) ds$.

For computational reasons, a discrete model is required. Let us represent the attenuation values by a vector $\mathbf{f} = [\mathbf{f}_{ijk}] \in \R^{N \times N \times T}$. Here $\mathbf{f}_{ijk}$ denotes the average of the values of the function $f$ over the voxel with indices $(i,j,k)$. 

The line integral can be approximated by
\begin{equation}
\int_L f(x) ds \approx \sum_{i=1}^N \sum_{j=1}^N\sum_{k=1}^T a_{ijk} \mathbf{f}_{ijk},
\end{equation}
where $a_{ijk}$ is a distance that the line $L$ travels in the voxel with indices $(i,j,k)$.
Then the practical three-dimensional tomographic X-ray data is modelled by
\begin{equation}
{\mathbf {m=Af + \varepsilon}},
\end{equation}
with a matrix $\mathbf{A}$ containing one row for each pixel in the set of measurements and an additive noise $\varepsilon$.

We use the normalized measurement matrix $\frac{{\mathbf A}}{\| {\mathbf A} \|}$ and measurement data $\frac{\mathbf{m}}{\| {\mathbf A} \|}$. 
Note that the norm of ${\mathbf A}$ equals $\sqrt{\lambda}$, where $\lambda$ is the largest eigenvalue of the symmetric matrix ${\mathbf A}^T{\mathbf A}$. The power method can be used to compute $\lambda$ in a matrix-free fashion \cite{golub2012matrix}.
Multiplication by the matrices ${\mathbf A}$ and ${\mathbf A}^T$ can be implemented by using the SPOT operator \cite{bleichrodt2016easy}.  

\subsection{The Shearlet Transform}

Shearlets form a directional representation system for multidimensional data \cite{guo2006sparse,labate2005sparse}. They can overcome limitations of traditional systems like wavelets that only provide optimally sparse representations for functions of one variable. In the 3D case, shearlets offer optimal approximation of piecewise smooth functions with jumps appearing only along smooth surfaces.

Shearlets are parameterized by scale, shearing and translation indices organized in the set
$$
\Lambda = \mathbb{N}_0 \times \{\lceil -2^{j/2} \rceil,\dots,\lceil2^{j/2} \rceil \} \times\mathbb{Z}^3.
$$
It has been shown in \cite{kutyniok2012} that, under suitable assumptions, the collection  $\psi_\gamma = \psi_{(l,m,n)} \in L^2(\R^3)$, where $\gamma = (l,m,n) \in \Lambda$, forms a {\it frame} for $L^2(\R^3)$ functions. The shearlet transform is defined as the following vector of coefficients:
\begin{equation}\label{model}
\mathcal{S}({\mathbf f})=
(\langle {\mathbf f},\psi_\gamma \rangle)_{\gamma\in\Lambda}.
\end{equation}
We use the ShearLab  implementation~\cite{kutyniok2016shearlab}.

\subsection{Sparsity-promoting regularization}

In this work, we are interested in finding the vector $\mathbf f$ that minimizes the variational regularization functional 
\begin{equation}\label{model}
\| {\mathbf A} {\mathbf f}-{\mathbf m}\|_2^2 + \mu \sum_{\gamma}|\langle {\mathbf f},\psi_\gamma \rangle|.
\end{equation} 
The parameter $\mu$ in (\ref{model}) describes a trade-off between emphasizing either the data fidelity term or the regularizing penalty term.

We introduce a regularization method based on enforcing sparsity in the shearlet transform domain. In their seminal paper~\cite{chen2016primal}, Peijun Chen, Jianguo Huang, and Xiaoqun Zhang show that the minimizer of (\ref{model}) can be computed using the primal-dual fixed point (PDFP) algorithm: 

\begin{equation}\label{PDFP}
\begin{split}
\y^{(i+1)} &= \mathbb{P}_C \Big(\f^{(i)} - \tau \nabla g(\f^{(i)}) -\lambda\SH^T \V^{(i)}\Big)\\
\V^{(i+1)} &= \Big(I - \mathcal{T}_\mu \Big) \Big(\SH \y^{(i+1)} + \V^{(i)} \Big)\\
\f^{(i+1)} &= \mathbb{P}_C \Big(\f^{(i)} - \tau \nabla g(\f^{(i)}) -\lambda\SH^T \V^{(i+1)} \Big)
\end{split}
\end{equation}
where  $\tau$ and $\lambda$ are positive parameters, $g(\f) = \frac{1}{2}\| \A \f- \m \|_2^2$, the matrix $\SH$ is a digital implementation of the shearlet transform and $\mathcal{T}$ is the
soft-thresholding operator defined by
\begin{equation}\label{eq:thresholding}
\mathcal{T}_\mu(c)=  
\begin{cases}
c + \frac{\mu}{2} &{\text {if }} x\leq -\frac{\mu}{2}\\
0 &{\text {if }} |x|<\frac{\mu}{2}\\
c - \frac{\mu}{2} &{\text {if }} x\geq -\frac{\mu}{2}.
\end{cases}
\end{equation}
Here $\mu > 0$ represents the thresholding parameter, while $\tau$ and $\lambda$ are parameters that need to be suitably chosen to guarantee convergence. In detail, $0 < \lambda < 1/ \lambda_{\max}(\SH\SH^T)$, where $\lambda_{\max}$ denotes the maximum eigenvalue, and $0 < \tau < 2/\tau_{\text{lip}}$, being $\tau_{\text{lip}}$ the Lipschitz constant for $g(\f)$. Furthermore, in (\ref{PDFP}) the non-negative ``quadrant'' is denoted by $C = \R_+^{N^2}$ and $\mathbb{P}_C$ is the Euclidian projection. In other words,  $\mathbb{P}_C$ replaces any negative elements in the input vector by zero.

\subsection{Automatic Selection of the  Threshold Parameter $\mu$}\label{Automatic Selection}

Assuming that we know {\it a priori} the expected degree of sparsity in the reconstruction, denoted $0<{\mathcal C}_{pr}\leq 1$. We use a simple feedback control system for finding such a value of $\mu$ that the iteration (\ref{PDFP}) produces a result with ${\mathcal C}_{pr}\cdot 100\%$ of its shearlet coefficients nonzero.

In our proposed method, $\mu=\mu^{(i)}$ is allowed to vary during the iteration. Furthermore, it is automatically tuned in every iteration to:
\begin{equation*}
\mu^{(i+1)} := \mu^{(i)} + \beta e^{(i+1)},
\end{equation*}
where $e^{(i+1)}={\mathcal C^{(i)}}-{\mathcal C}_{pr}$ and $0\leq{\mathcal C}^{(i)}\leq 1$ is the sparsity level of the current iterate $\mathbf{f}^{(i)}$. If $\beta > 0$ is too large, the controller induces oscillations in the regularization parameter and if it is too small convergence is slow. To avoid this, we choose a large initial value for beta, but decrease it each time the sparsity level crosesses the desired level of sparsity: %To avoid slow convergence, the stepsize $\beta>0$ cannot be too small and to address this we choose large beta but decrease it each time the sparsity level crosses the desired level of sparsity:
\begin{equation*}
\beta = \beta (1 - |e^{(i+1)} - e^{(i)}|).
\end{equation*}
This approach can avoid unwanted oscillation in the values of $\mu^{(i)}$.

\subsection{Bone Quality Measures}\label{Subsec Quality Measures}

We study human trabecular bone samples. The recommended parameters for studying the 3D structure of bone \cite{bouxsein2010guidelines,parfitt1987bone,marinozzi2012variability} include:
\begin{enumerate}
\item Percentage of bone volume ({\it BV/TV}). BV refers to volume of the region segmented as bone and {\it BV/TV} refers to the ratio of the segmented bone volume to the total volume of the volume of interest (VOI); 
\item Trabecular thickness ({\it Tb.Th}): the diameter of the largest sphere which is entirely bounded within the solid surfaces (mm);
%\item Trabeculae number ({\it Tb.N}): measure of the average number of trabecular per unit length ($\frac{1}{\mu m}$)
\item Trabecular separation ({\it Tb.Sp}): the thickness of the spaces as defined by binarization within the VOI (mm).
\end{enumerate}

To calculate the basic bone morphometric parameters, standard Computed Tomography Analyzer (CTAn) software provided by the manufacturer (Bruker microCT, Kontich, Belgium) was used. The reconstructed images were converted to 8-bit images and then segmented into binary images for morphometric analysis in CTAn. Because the samples were drilled from bone, physical artifacts such as bone dust or cracks were left from the preparation. Therefore, a volume of interest (VOI) inside the sample was selected so that the edge artifacts would not affect the analysis.

\subsection{Determining the {\it A Priori} Degree of Sparsity}\label{Selection of sparsity}
Denote $\f^\kappa$ as the best $\kappa-$term shearlet approximation using the $\kappa$ largest coefficients in the shearlets expansion \cite{devore1998nonlinear}.  
We compute the best $\kappa-$term approximations of the baseline (FDK reconstruction from full projection images) image using different values of $\kappa$. Once we computed the images $\{\f^\kappa\}$:
\begin{enumerate}
\item the morphometric parameters of trabecular bone ({\it BV/TV, Tb.Th} and {\it Tb.Sp}) for each images are computed and
\item the plate thickness value for the standard phantom for each images are computed.
\end{enumerate}

At particular level, as the sparsity level $\kappa$ decreases, the morphometric parameters and the plate thickness value start to deteriorate. The prior sparsity level, $C_{pr}$ is chosen at the stage before at least one of the parameters in 1) and 2) for $\{\f^\kappa\}$ start to deteriorate.

\subsection{Pseudo-algorithm}\label{PseudoAlgorithm}

A step-by-step description of the proposed CSDS 
algorithm is summarized in Algorithm \ref{alg:Algorithm1}. As an addition, given $\mu\ge 0$, for a vector $s\in\R^{N^2\times T \times K}$, where $K$ is the number of 3D shearlets, we define the number of elements larger than $\mu$ in absolute value as follows: 
\begin{equation}\label{bestApprox}
\#_\mu s := \#\{\,i\,\,| 1\leq i\leq N^2\times T \times K,\  |s_i|>\mu\}.
\end{equation}
% Now, the sparsity level is defined by
% \begin{equation}
% {\mathcal{C}} =  \frac{\#_\kappa\{\mathbf S \f\}}{N^2\times T \times K},
% \end{equation}
% where $N^2\times T \times K$ is the total number of shearlets coefficients. In practical computations the value of $\kappa$ is set to be small but positive.

\begin{algorithm}
\caption{Shearlet sparsity-promoting tomographic reconstruction algorithm with the automatic parameter choice method. Inputs: measurement data vector $\mathbf{m}$, system matrix $\mathbf{A}$, parameters $\tau, \, \lambda > 0$ to ensure convergence,
{\it a priori} degree of sparsity ${\mathcal C}_{pr}$,
initial thresholding parameter $\mu^{(0)}$,
the maximum number of iterations $I_{\max}>0$, 
tolerances $\epsilon_1,\epsilon_2>0$ for the stopping rule and control stepsize $\beta>0$.}
\begin{algorithmic}[1]
\State Inputs: measurement data vector $\mathbf{m}$, system matrix $\mathbf{A}$, parameters $\tau, \, \lambda > 0$ to ensure convergence,
{\it a priori} degree of sparsity ${\mathcal C}_{pr}$,
initial thresholding parameter $\mu^{(0)}$,
the maximum number of iterations $I_{\max}>0$, 
tolerances $\epsilon_1,\epsilon_2>0$ for the stopping rule and
control stepsize $\beta>0$.

\State $\f^{(0)} = \mathbf{0}$, $i=0$, $e=1$, and $\mathcal{C}^{(0)} = 1$ 
\While{$i<I_{\max}$ and $|e|\geq\epsilon_1$ or $d \geq \epsilon_2$}
\State $e=\mathcal{C}^{(i)}-\mathcal{C}_{pr}$
\If {$\sign(e^{(i+1)}) \neq \sign(e^{(i)})$}
\State $\beta = \beta (1 - |e^{(i+1)} - e^{(i)}|)$ 
\EndIf
\State $\mu^{(i+1)} = \max\{0, \mu^{(i)} + \beta e \}$
\State $\y ^{(i+1)} = \max \{0,\f^{(i)} - \gamma \nabla g_1(\f^{(i)})-\lambda\mathbf{S}^T \V^{(i)}\} $
\State $\V^{(i+1)} = (I - \mathcal{T}_{\mu^{(i)}})(\mathbf{S} \y^{(i+1)} + \V^{(i)})$
\State $\f^{(i+1)} = \max \{0,\f^{(i)}{-}\gamma \nabla g(\f^{(i)}){-}\lambda\mathbf{S}^T \V^{(i+1)}\}$
\State ${\mathcal C^{(i+1)}} = \#_{\mu^{(i+1)}}(\mathbf{S} \f^{(i+1)})/N^{2}TK$
\State $d = \Vert \f^{(i+1)} - \f^{(i)} \Vert_2 / \Vert \f^{(i+1)} \Vert_2$
\State $i := i+1$
\EndWhile
\end{algorithmic}
\label{alg:Algorithm1}
\end{algorithm}

\section{Experiments}

\subsection{Data Acquisition}\label{Data Acquisition}
%\subsubsection{Phantom}

\subsubsection{Human trabecular bone}

X-ray data from two osteochondral samples  were acquired. Two samples (diameter = 4 mm) were harvested from the weight bearing area of tibial plateus from two cadavers under the approval of The Research Ethics Committee of the Northern Savo Hospital District, Kuopio, Finland (approval no 134/2015). The X-ray tomography data was acquired with a SkyScan 1272 high-resolution $\mu$CT scanner (Bruker microCT, Kontich, Belgium). The isotropic voxel size length for projection 22 $\mu$m/pixel and the number of frames averaged is 2 per projection. The TIFF files are the input to compute the FDK and CSDS algorithms and their dimensions are $1008 \times 672$. We collected $300$ projection images acquired over a full 180 degree rotation with uniform angular step of 0.6 degrees between projections. Each projection image was composed of 1500 ms exposures. The X-ray tube acceleration voltage was 50 kV and the tube current 200 $\mu$A. The full polychromatic beam was used for image acquisition. The additional filtration was 0.5 mm of Aluminium.
%{\color{red}(Only my question:why and what the filter do?)}.

We used 300 complete projections for baseline reconstructions. We picked two subsets of projections (30 and 50) from the measured data with uniform angular sampling from different total opening angles of each projection image. 
% See Table~\ref{Opening angles}.
% \begin{table}
% \caption {Total angle of view for each restricted set of projection images} \label{Opening angles}
% \begin{center} 
% \begin{tabular}{ | c | c |  c | } \hline 
% Number of projections & Total angle of view   \\ \hline 
% \hline
% 120    &  $178.5^{\circ}$ \\
% 60    &   $177.0^{\circ}$ \\
% 40   &   $175.5^{\circ}$\\
% \hline
% \end{tabular} 
% \end{center}
% \end{table}

\subsection{Morphometrics Parameters Values using Different Sparsity Levels}
The baseline $\f_{pr}$ was computed using the FDK method with the full set of $300$ projection images. The reconstruction images for both trabecular samples have a size of $240\times 240 \times 180$.

\begin{figure}
\begin{picture}(200,100)
%\centering
\put(-10,0){\includegraphics[width=5cm]{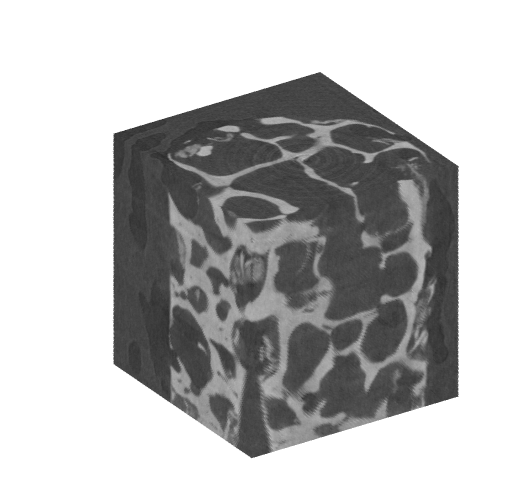}}
\put(140,){\includegraphics[width=5cm]{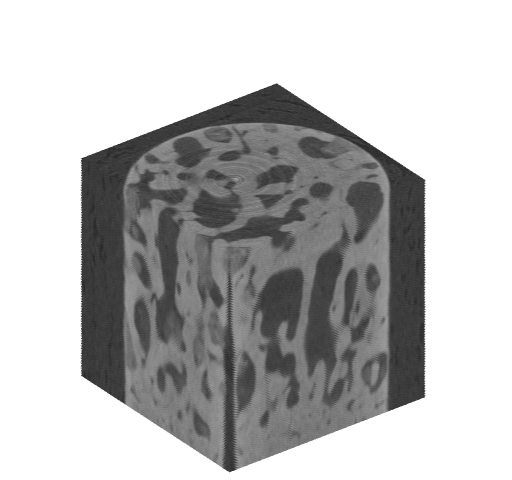}}
\put(55,-5){(a)}
\put(195,-5){(b)}
\end{picture}
\caption{\label{FullBone} 3D reconstructions of healthy (a) and osteoarthritis (b) human trabecular bone using FDK method from 300 projections.}
\end{figure}

\begin{figure}
 \begin{picture}(88,210)
 \put(13,145){\includegraphics[width=8cm]{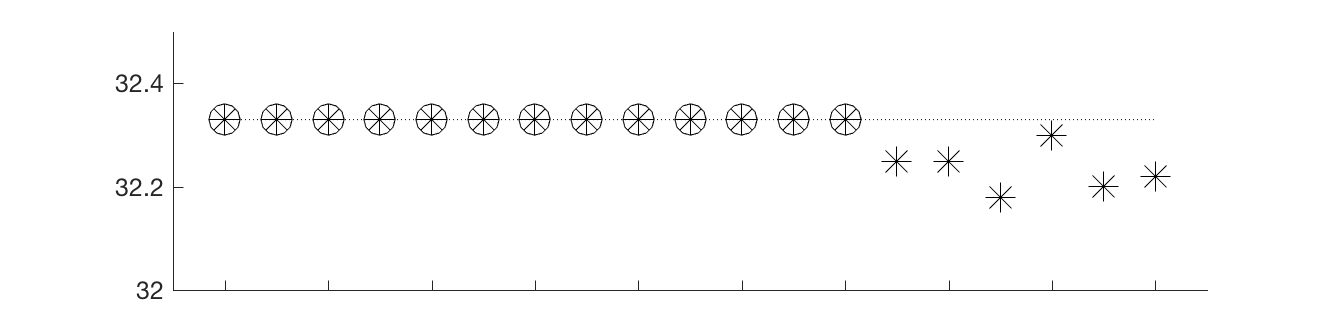}}
 \put(13,75){\includegraphics[width=8cm]{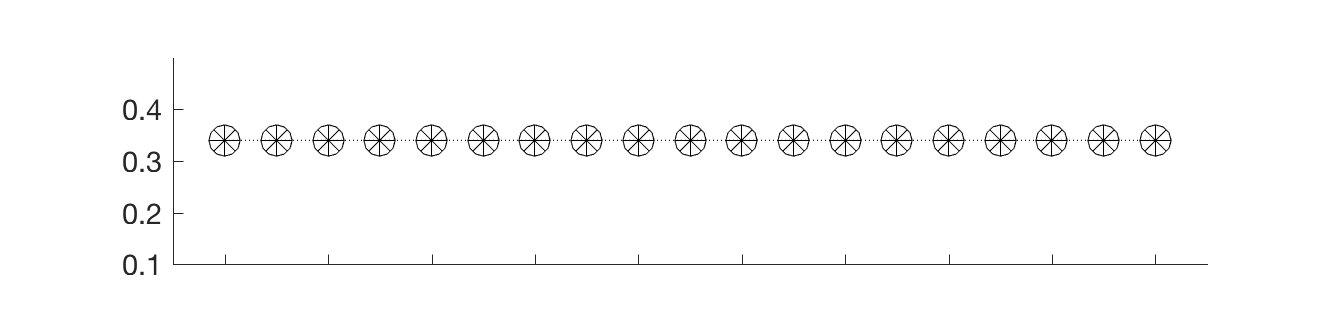}}
 \put(13,0){\includegraphics[width=8cm]{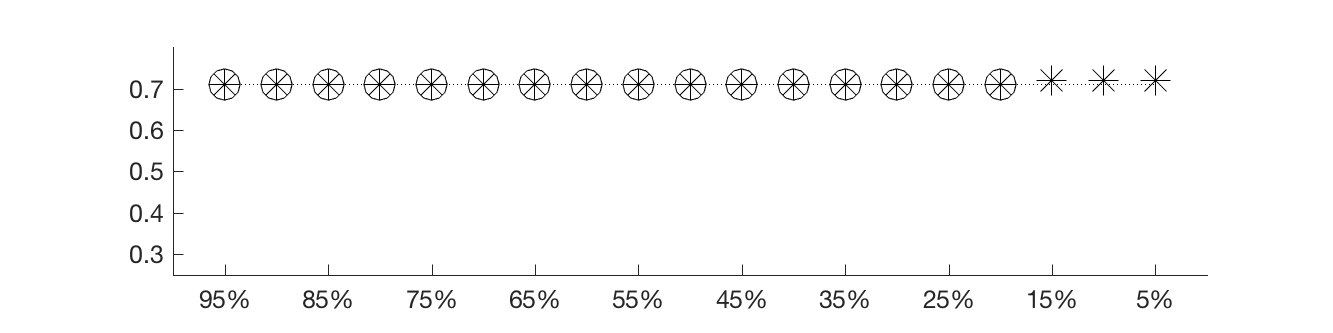}}
 \put(40,203){\it BV/TV}
 \put(40,135){\it Tb.Th}
 \put(40,62){\it Tb.Sp}
 \end{picture}
% \caption{\label{fig:HealthySparsity} The value of trabecular bone parameters from inverse 3D shearlet transform of the healthy bone using different sparsity level as discussed in the step above. Stars with circles are the stable values.}
  \caption{\label{fig:HealthySparsity} The value of trabecular bone parameters from the best $\kappa$-term approximation images of the healthy bone using different sparsity level $\kappa$ as discussed in \ref{Selection of sparsity}. Stars with circles are the stable values.}
 \end{figure}

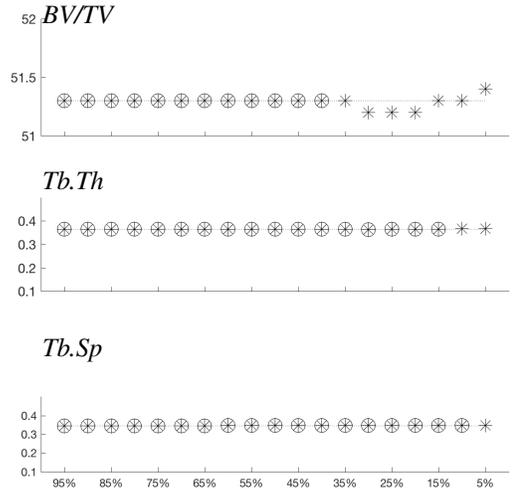
\begin{figure}
 \begin{picture}(88,210)
 \put(13,135){\includegraphics[width=8cm]{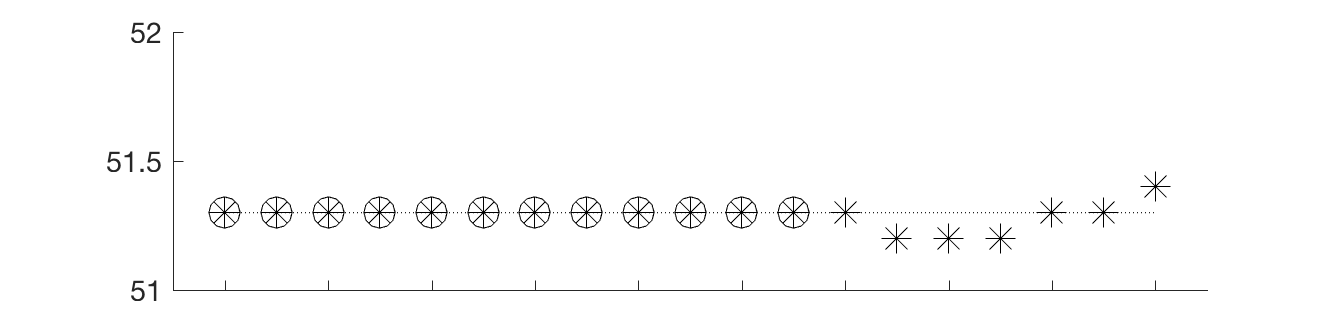}}
 \put(13,72){\includegraphics[width=8cm]{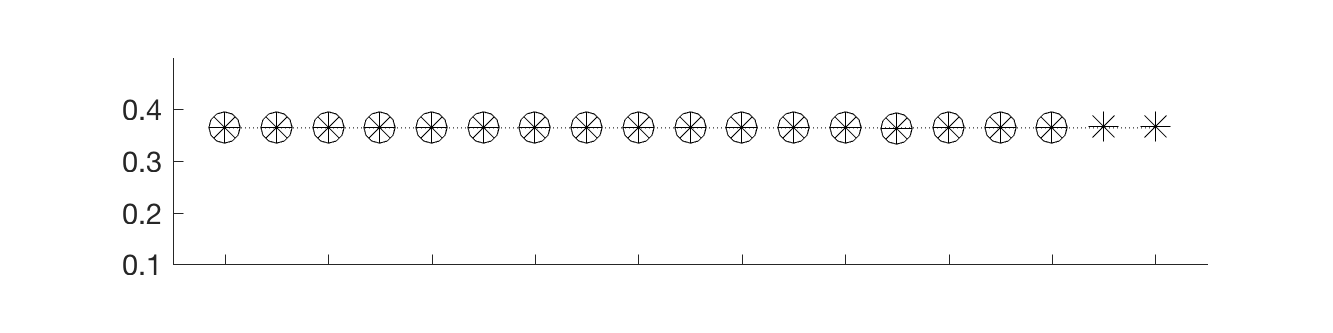}}
 \put(13,0){\includegraphics[width=8cm]{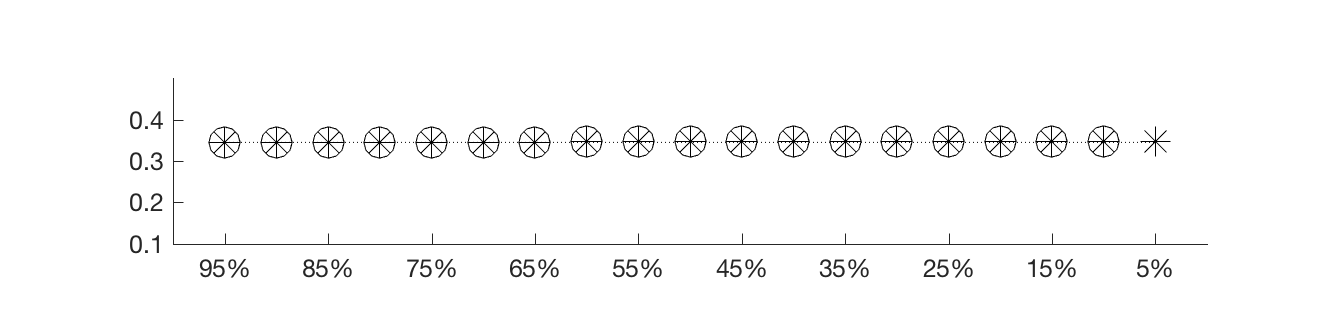}}
 \put(43,185){\it BV/TV}
 \put(43,122){\it Tb.Th}
 \put(43,59){\it Tb.Sp}
 \end{picture}
 \caption{\label{fig:OsteoSparsity} The value of trabecular bone parameters from the best $\kappa$-term approximation images of the osteoarthritic bone using different sparsity level $\kappa$ as discussed in \ref{Selection of sparsity}. Stars with circles are the stable values.}
% \caption{\label{fig:OsteoSparsity} The value of trabecular bone parameters from inverse 3D shearlet transform of the osteoarthritic bone using different sparsity level as discussed in the step above. Stars with circles are the stable values.}
 \end{figure}

A collection of best $\kappa-$term approximation images, $\f_{pr}^\kappa$ were computed using the strategy discussed in Subsection~\ref{Selection of sparsity}. This was done for the bone samples (healthy and OA) as it is shown in Figure~\ref{FullBone}. The thresholded parameters $\kappa$ were selected from $95\%$ to $5\%$. The trabecular morphometrics parameters and plate thickness parameter of the phantom were calculated for each $\kappa$.
% The calculations of the morphometric parameters of the healthy and the OA trabecular bone can be seen in Figures~\ref{fig:HealthySparsity} and \ref{fig:OsteoSparsity}, respectively. We can see {\it BV/TV} starts to deteriote at $\kappa = 40\%$. 
%So, we can choose the {\it a priori} degree of the sparsity level as the $35\% - 40\%$ largest shearlet coefficients. 
For $\kappa=35\%$ in the healthy sample and $\kappa=40\%$ in the OA sample, {\it BV/TV} parameter of the trabecular bone starts to deteriorate as shown in Figures~\ref{fig:HealthySparsity} and \ref{fig:OsteoSparsity}, respectively. We choose the {\it a priori} degree of the sparsity level as the mean of both $\kappa$s.  We denote the {\it a priori} degree of the sparsity level as ${\mathcal C}_{pr}$. 

\section{Results}
\noindent
%For the reconstruction, we set a 3D image sized $N \times N \times T$ with $N = 256$ and $T = 100$. 
The same size of reconstruction images were set for FDK and CSDS algorithms using a different number of projection images. All the algorithms were implemented in Matlab. For FDK reconstructions, the experiments were performed on Intel(R) Xeon(R) CPU E5-1650 v3 at 3.7GHz RAM 32 and GPU 4GB memory. The ASTRA Toolbox (iMinds-Vision Lab, University of Antwerp, Belgium) and Spot operator were used in reconstructions \cite{bleichrodt2016easy,van2016fast,van2015astra}. The CSDS computations was performed on CPU at supercluster taito.csc.fi. All of the computations were set up using a cone beam geometry. 
%The collection of projection images was downsampled with $3 \times 3$ binning to overcome computational challenges. For the sake of visualization, only one slice of 3D-reconstructions using shearlet-based method and FDK are presented. 

%In details, in the shearlet transform, the scale of transformation is set to 2.  
The details are as follows: we used ${\mathcal C}_{pr} = 37.5\%$, the initial value for the thresholding parameter $\mu^0$ was calculated from the absolute mean of $(1-\mathcal{C}_{pr})$ of the shearlet coefficients from the backprojection reconstruction. The number of scale 3D shearlet transform is set equal to 1. The maximum number of iteration $I_0 = 1000$ is set for additional termination criteria. However, in the computation, the number of iteration never reached the $I_0$.

\begin{figure*}
\centering
\begin{picture}(380,250)
\put(-60,115){\includegraphics[width=3.75cm]{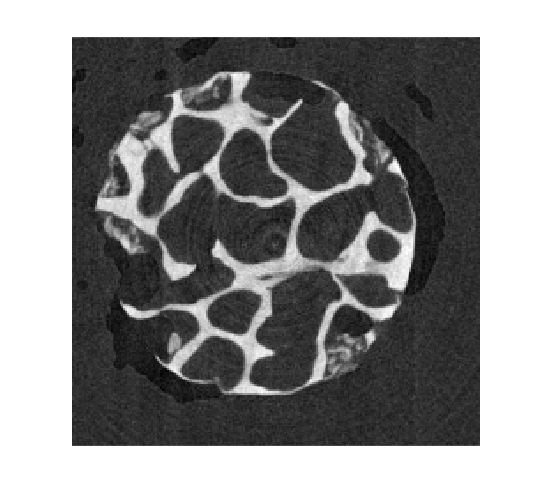}}
\put(40,115){\includegraphics[width=3.75cm]{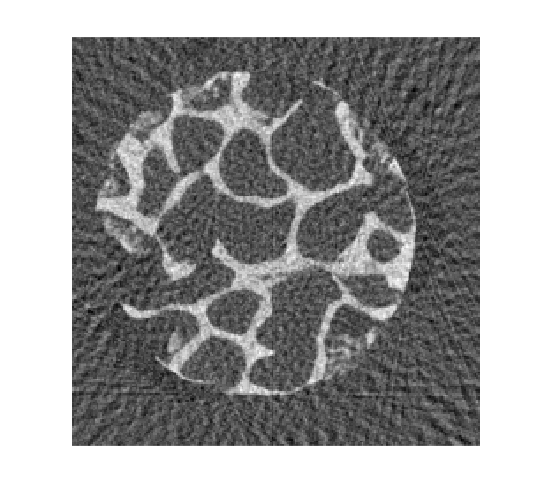}}
\put(140,115){\includegraphics[width=3.75cm]{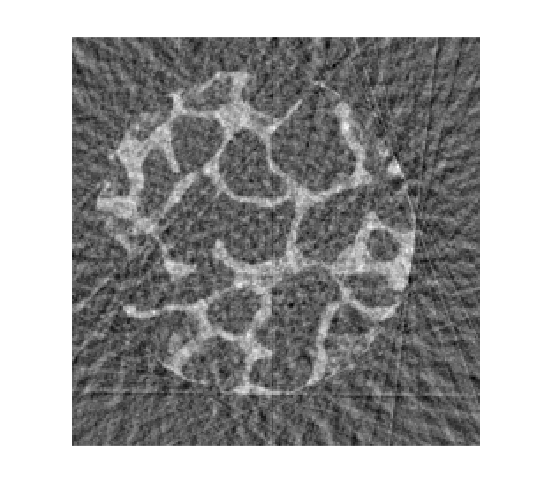}}
\put(240,115){\includegraphics[width=3.75cm]{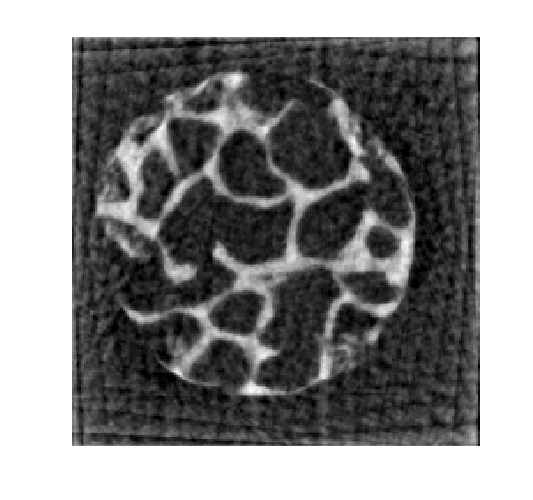}}
\put(340,115){\includegraphics[width=3.75cm]{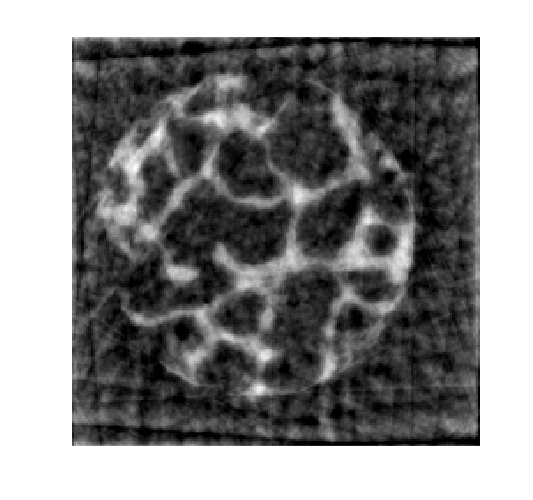}}
\put(-60,0){\includegraphics[width=3.75cm]{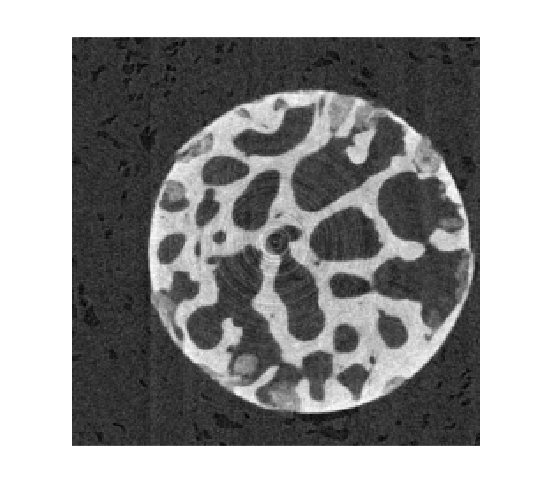}}
\put(40,0){\includegraphics[width=3.75cm]{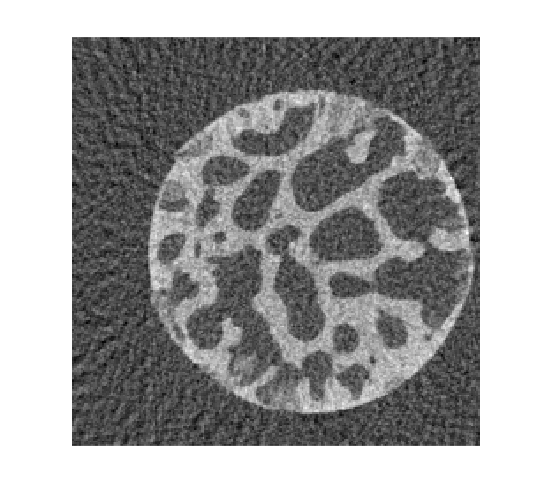}}
\put(140,0){\includegraphics[width=3.75cm]{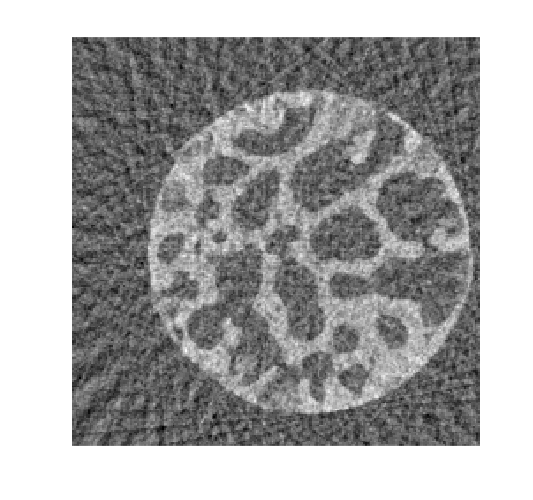}}
\put(240,0){\includegraphics[width=3.75cm]{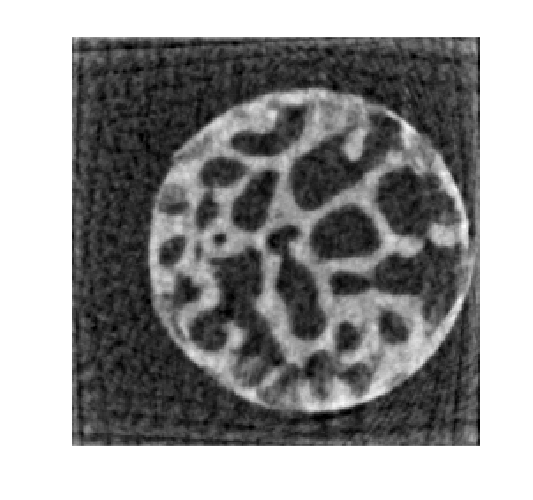}}
\put(340,0){\includegraphics[width=3.75cm]{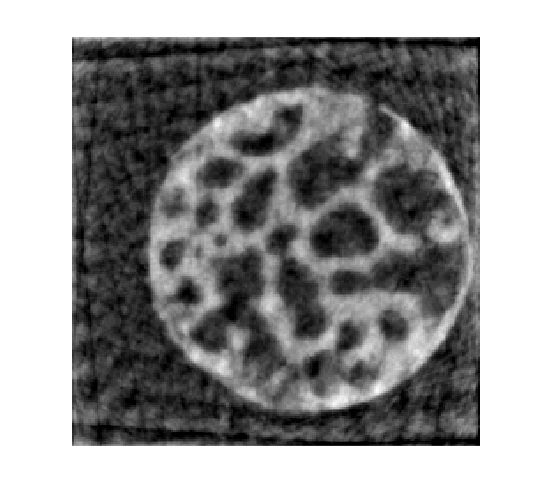}}
\put(-57,60){\raisebox{-1.5\normalbaselineskip}[0pt][0pt]{\rotatebox{90}{OA}}}
\put(-57,160){\raisebox{-1.5\normalbaselineskip}[0pt][0pt]{\rotatebox{90}{healthy}}}
\put(-12,110){(a)}
\put(89,110){(b)}
\put(189,110){(c)}
\put(290,110){(d)}
\put(390,110){(e)}
\put(-12,-5){(f)}
\put(89,-5){(g)}
\put(189,-5){(h)}
\put(290,-5){(i)}
\put(390,-5){(j)}
\end{picture}
\caption{\label{fig:reconstructions}Axial micro-CT cross-section images of the 3D reconsructions. The baseline images (FDK reconstruction from 300 projections) are given in (a) and (f), FDK reconstructions from $50$ projections are shown in (b) and (g), FDK  reconstructions from $30$ projections are shown in (c) and (h),  CSDS  reconstructions from $50$ projections are shown in (d) and (i) and CSDS reconstructions from $30$ projections are shown in (e) and (j).}
\end{figure*}
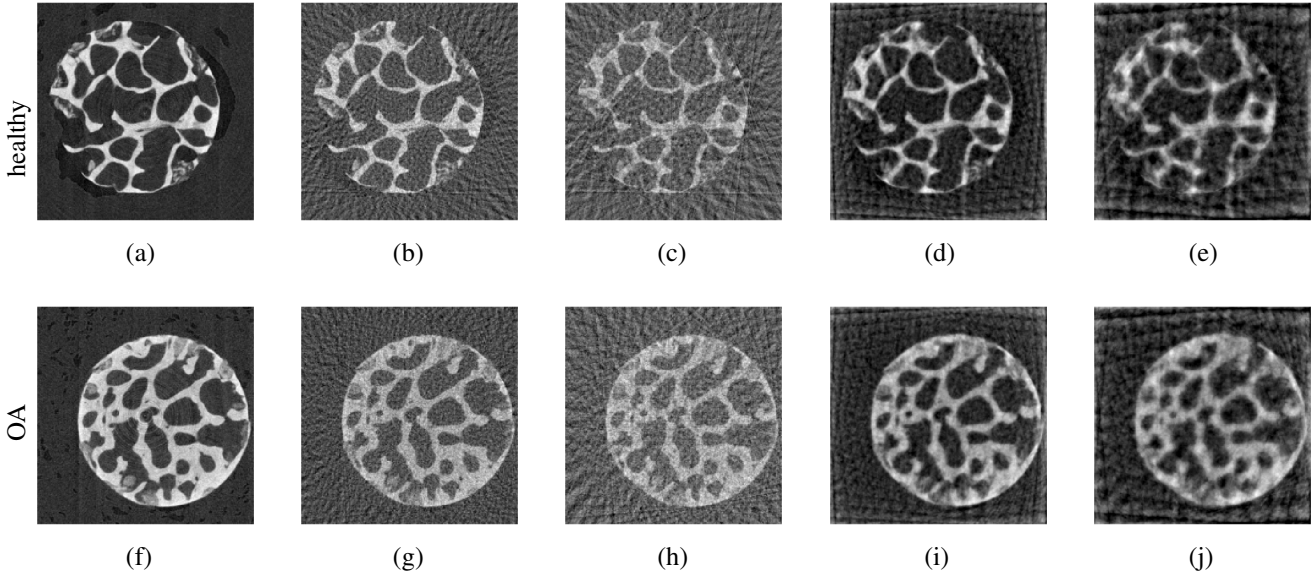

\begin{figure*}
\centering
\begin{picture}(405,250)
\put(-10,115){\includegraphics[width=2.5cm]{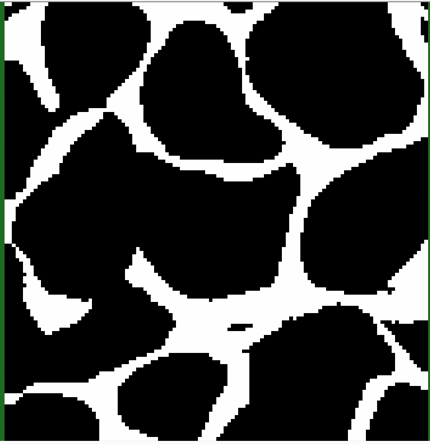}}
\put(80,115){\includegraphics[width=2.5cm]{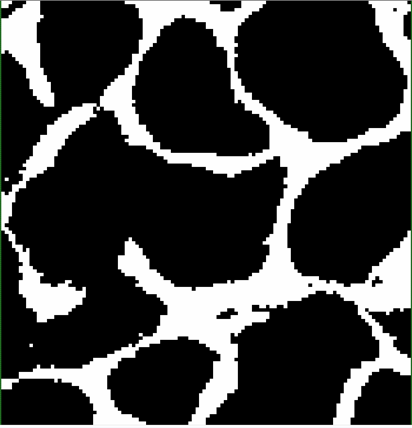}}
\put(170,115){\includegraphics[width=2.5cm]{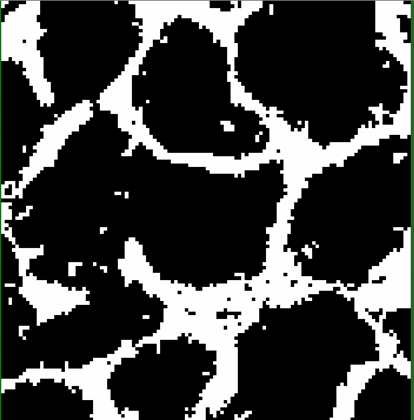}}
\put(260,115){\includegraphics[width=2.5cm]{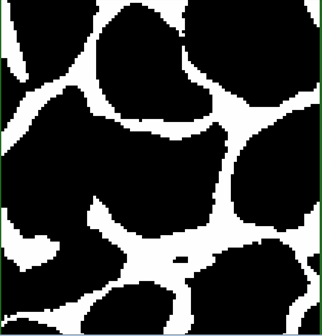}}
\put(350,115){\includegraphics[width=2.5cm]{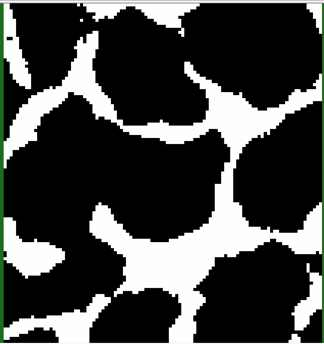}}
\put(-10,10){\includegraphics[width=2.5cm]{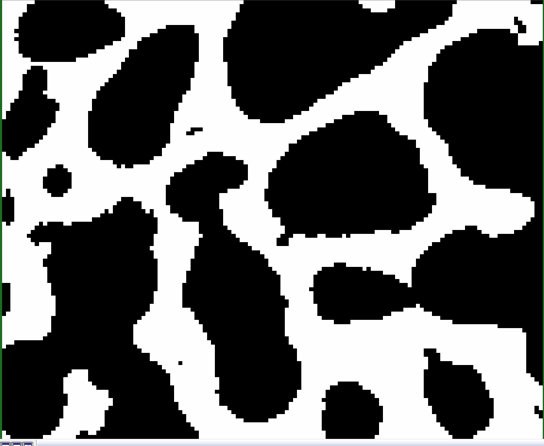}}
\put(80,10){\includegraphics[width=2.5cm]{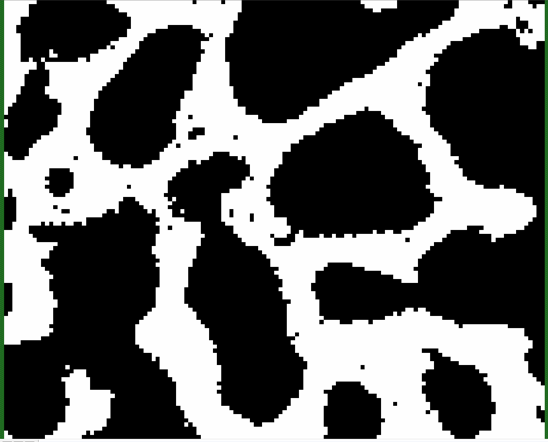}}
\put(170,10){\includegraphics[width=2.5cm]{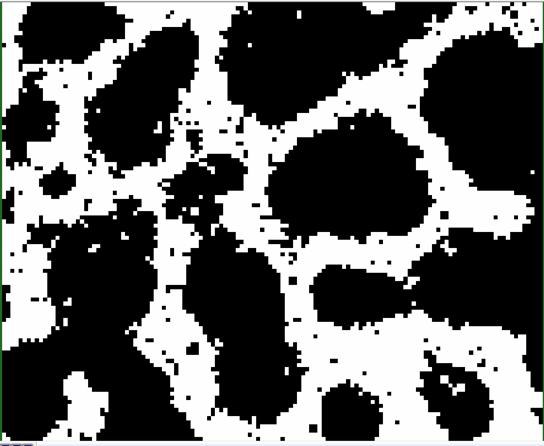}}
\put(260,10){\includegraphics[width=2.5cm]{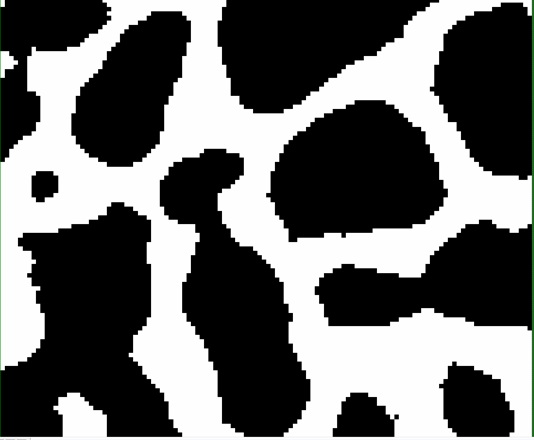}}
\put(350,10){\includegraphics[width=2.5cm]{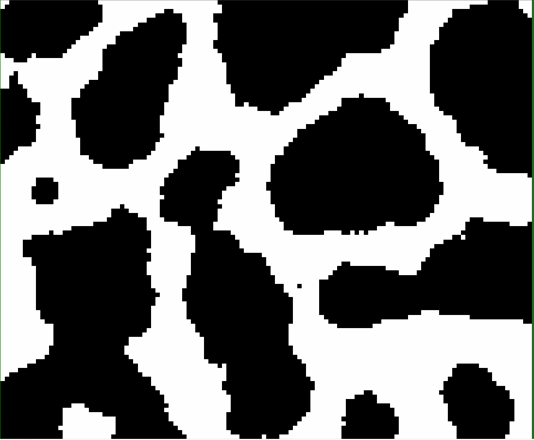}}
\put(-25,155){\raisebox{-1.5\normalbaselineskip}[0pt][0pt]{\rotatebox{90}{healthy}}}
\put(-25,55){\raisebox{-1.5\normalbaselineskip}[0pt][0pt]{\rotatebox{90}{OA}}}
\put(20,100){(a)}
\put(110,100){(b)}
\put(200,100){(c)}
\put(290,100){(d)}
\put(380,100){(e)}
\put(20,-5){(f)}
\put(110,-5){(g)}
\put(200,-5){(h)}
\put(290,-5){(i)}
\put(380,-5){(j)}
\end{picture}
\caption{\label{fig:binary images}Axial cross-section images of thresholded images or binary images of the volume of interest which correspond to Figure~\ref{fig:reconstructions}.}
\end{figure*}
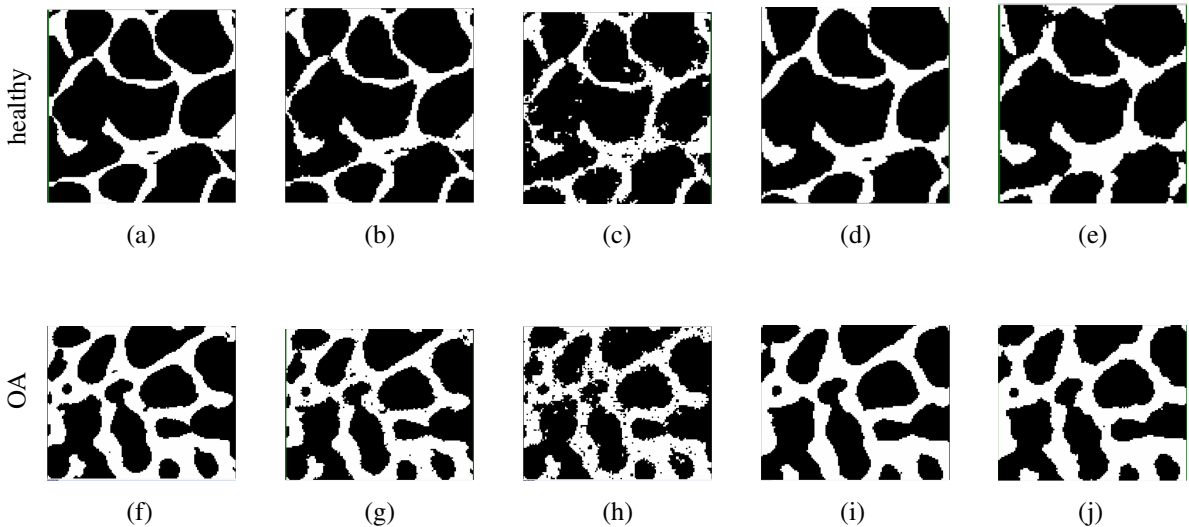

We set the control step size $\beta = 10\mu^0$, $\epsilon_1 = 5\times 10^{-3}$ and $\epsilon_2 = 1\times 10^{-3}$ as the stopping rule. The shearlet-based reconstructions are shown in Figure~\ref{fig:reconstructions}. 
The ROI of the reconstruction images were chosen and segmented by applying the steps in Subsection \ref{Subsec Quality Measures}. The segmented images are shown in Figure~\ref{fig:binary images}. 

For comparison, FDK reconstructions of healthy and OA trabeculae were computed as well as can be seen in Figure~\ref{fig:reconstructions}. The thresholded images of FDK reconstructions are presented in Figure~\ref{fig:binary images}. In addition, the trabecular bone morphometrics parameters for FDK and shearlet-based reconstructions were calculated and given in Table~\ref{healthy parameter} and Table~\ref{OA parameter}.

\begin{figure}
%\centering
\begin{picture}(0,300)
\put(-7,275){\includegraphics[width=8cm]{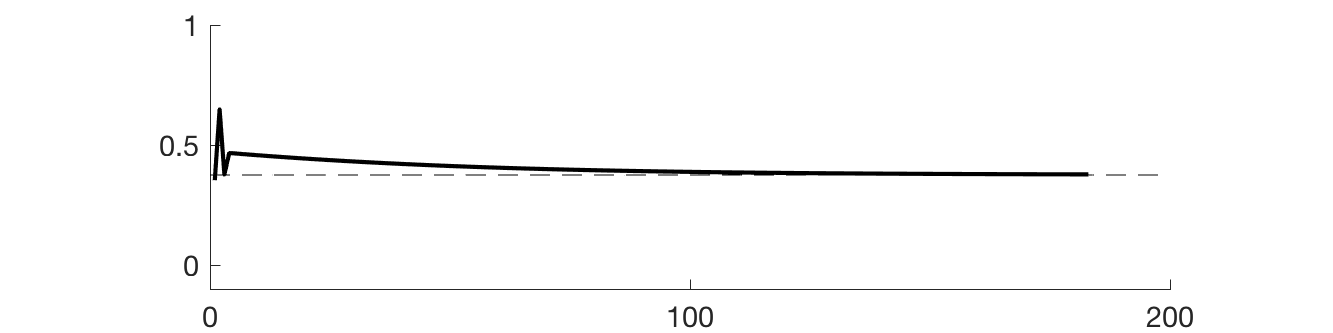}}
\put(0,185){\includegraphics[width=8cm]{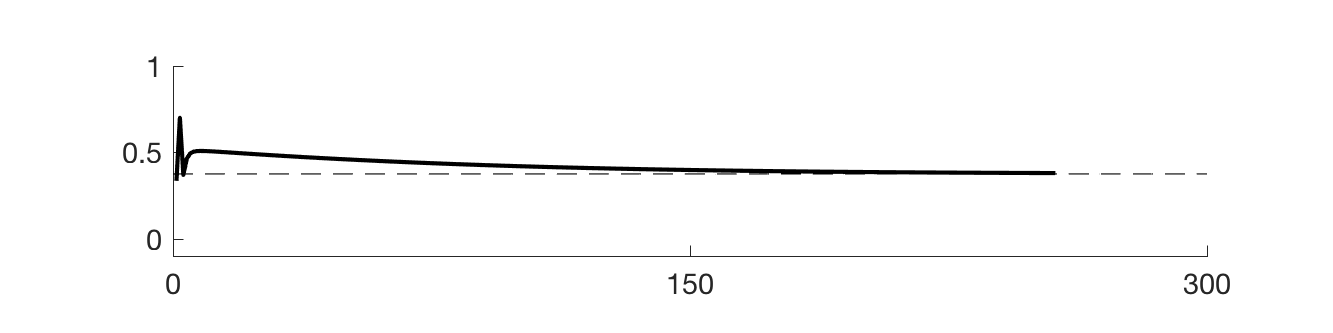}}
\put(0,95){\includegraphics[width=8cm]{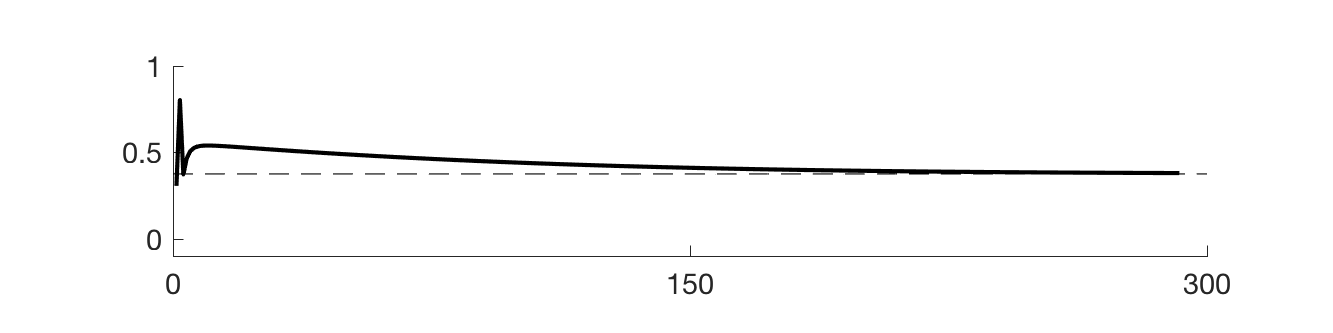}}
\put(-7,5){\includegraphics[width=8cm]{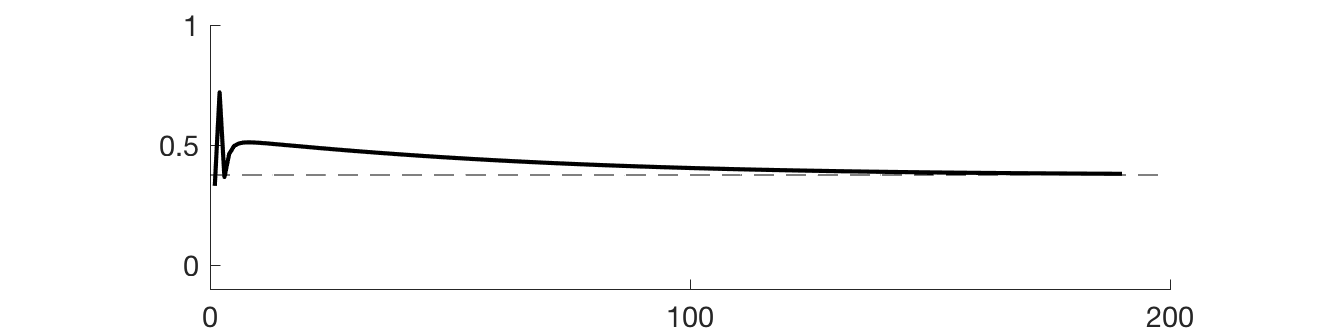}}
\put(110,-5){(d)}
\put(110,75){(c)}
\put(110,170){(b)}
\put(110,260){(a)}
\end{picture}
\caption{\label{fig:sparvec} The ratio of nonzero shearlet coefficients as the iteration
progresses (thick line). The dashed line is the $C_{pr}$ (a) and (b): the healthy sample using 30 and 50 projections. (c) and (d): the OA sample using 30 and 50 projections}
\end{figure}
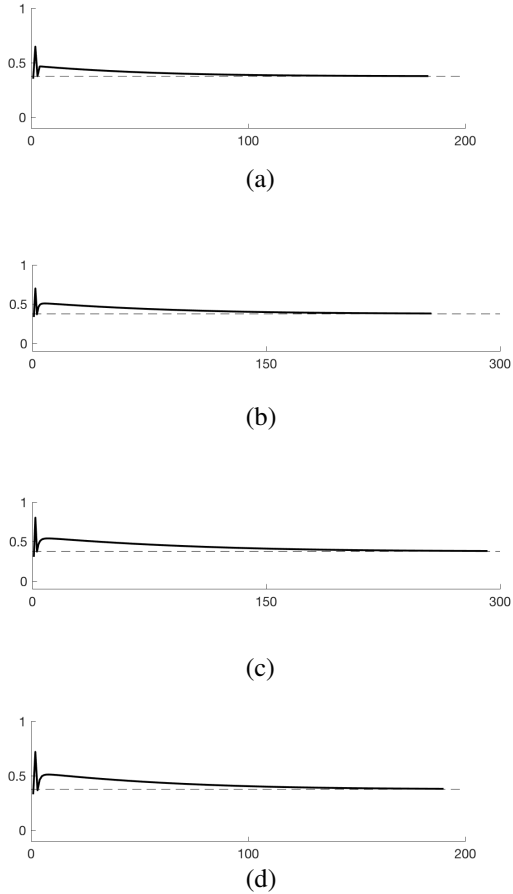

\begin{table}[h]
\caption {The trabecular bone morphometric parameters calculation for the healthy sample reconstruction from different number of projection images.} \label{healthy parameter}
\begin{center} 
\begin{tabular}{ | r | c| c |  c | c | } \hline 
Method & Number of  &  {\it BV/TV}  & {\it Tb.Th}  & {\it Tb.Sp}   \\ 
&projections&&(mm)&(mm)\\
\hline 
\hline
Baseline& 300  &  32.33\% & 0.34 & 0.71   \\  
\hline
FDK& 50   &  30.50\% & 0.28 & 0.64\\ 
&30   &  32.17\% & 0.24& 0.51  \\ 
\hline 
CSDS& 50   & 33.77\% & 0.36 & 0.70  \\ 
&30   &  34.29\% & 0.33 & 0.63  \\ 
\hline  
\end{tabular} 
\end{center}
\end{table}

\begin{table}[h]
\caption {The trabecular bone morphometric parameters calculation for the OA sample reconstruction from different numbers of projection images.} \label{OA parameter}
\begin{center} 
\begin{tabular}{ | r | c| c |  c | c | } \hline 
Method & Number of  &  {\it BV/TV}  & {\it Tb.Th}  & {\it Tb.Sp}   \\ 
&projections&&(mm)&(mm)\\
\hline 
\hline
Baseline& 300  &  51.30\% & 0.37 & 0.35   \\  
\hline
FDK& 50   &  50.60\% & 0.30 & 0.29\\ 
&30   &  48.57\% & 0.21 & 0.21 \\ 
\hline 
CSDS& 50   & 53.69\% & 0.36 & 0.31 \\ 
&30   &  52.79\% & 0.37 & 0.33 \\ 
\hline  
\end{tabular} 
\end{center}
\end{table}

%\begin{table}
%\caption {The plate thickness and porosity calculation for the healthy sample reconstruction from different number of projection images.} \label{healthy porosity}
%\begin{center} 
%\begin{tabular}{ | r | c| c |  c |  } \hline 
%Method & Number of  &  {\it Pl.Th}  & Porosity \\ 
%&projections&&\\
%\hline 
%\hline
%Baseline& 300  &  272.5 & 2.21   \\  
%\hline
%FDK& 50   &  271.8 & 1.67 \\ 
%&30   &  270.0 & 1.33\\ 
%\hline 
%CSDS& 50   & 278.9 & 1.51  \\ 
%&30   &  271.0 & 2.20   \\ 
%\hline  
%\end{tabular} 
%\end{center}
%\end{table}

%\begin{table}
%\caption {The plate thickness and porosity calculation for the OA sample reconstruction from different number of projection images.} \label{healthy porosity}
%\begin{center} 
%\begin{tabular}{ | r | c| c |  c |  } \hline 
%Method & Number of  &  {\it Pl.Th}  & Porosity \\ 
%&projections&&\\
%\hline 
%\hline
%Baseline& 300  &  348.20 & 1.75   \\  
%\hline
%FDK& 50   &  354.83 & 0.14 \\ 
%&30   &  350.32 & 0.23\\ 
%\hline 
%CSDS& 50   & 368.56 & 0.47 \\ 
%&30   &  351.62 & 0.39   \\ 
%\hline  
%\end{tabular} 
%\end{center}
%\end{table}

\begin{table}
\caption {Computation times of 3D reconstruction using FDK method (in seconds)} \label{FDK Computation time}
\begin{center} 
\begin{tabular}{ | c | c |  c | c | } \hline 
Number of projections &  FDK for healthy and OA samples   \\ \hline 
\hline
%120    & 4.4  \\
50    & 5.0   \\
30   & 3.0   \\
\hline
\end{tabular} 
\end{center}
\end{table}

\begin{table}
\caption {Computation times of 3D reconstruction using shearlet-based method for healthy and OA samples (in seconds)} \label{Shearlet Computation time}
\begin{center} 
\begin{tabular}{ | c | c |  c | c | } \hline 
Number of &  shearlet-based method   & shearlet -based method\\ 
projections images & for healthy bone & for OA bone \\ \hline
\hline
50    &64\,001 & 60\,768   \\
& (256 iterations)&(292 iterations)\\
30   &46\,642 & 47\,715   \\
& (183 iterations) &(190 iterations)\\
\hline
\end{tabular} 
\end{center}
\end{table}

\section{Discussion and Conclusion}

We have presented X-ray reconstructions of the inner structures of the healthy and osteoarthritic human trabecular bone in 3D using sparse projection images. The use of limited data is beneficial in reducing the often long scan times and avoiding  massive amounts of data. Another advantage is to avoid moving artefacts.
%However, in this study {\it in vitro} samples were used. 

While traditional methods such as FDK require dense projection images to produce good reconstructions, we propose the shearlet-based method with automatically chosen regularization parameter for robust reconstruction for the incompletely sampled datasets. When the number of projection images was reduced, the significant streak artefacts overwhelm the FDK reconstruction images while the CSDS reconstructions contain less streak artefacts. The non-negativity constraint and the enforcement of the penalty term $\ell_1$-norm combined with the sparsity transform which acts as a denoising process in the CSDS method give significant contribution to produce better reconstructions as is shown in Figure~\ref{fig:reconstructions}.  

Figure~\ref{fig:sparvec} presents the behaviors of the sparsity levels for each datasets. In the initial iterations, short oscillations appear due to the large value of the tuning parameter $\beta$. They soon disappear as the `crossing' checking process decreases the parameter, as discussed in Section~\ref{Automatic Selection} and eventually the ratio of nonzero coefficients, ${\mathcal C}_{i}$ converges to ${\mathcal C}_{pr}$. This is one of the benefits of the CSDS proposed method, as the manual tuning could be avoided.

Besides the visual inspection, in this particular problem, we also measured the quality of the reconstructions quantitatively. We computed the morphometric parameters of the reconstructions, and compare them to the parameters from the baseline images.

The computations of morphometric parameters were done by using the standard steps in \ref{Subsec Quality Measures}. In the FDK reconstruction from undersampled data, the quality of the binary images were relatively poor since many speckles induced by the noise randomly appeared in the binary images. It can be seen in Figure~\ref{fig:binary images} that many of trabeculae were also broken. As a result, trabecular bone morphometric parameters progressively showed considerable differences when fewer projection images was used compared to the baseline (full projection images). For instance, for both samples, the FDK reconstructions using $50$ projections had differences in the {\it Tb.Sp} parameter of $9.86\% - 17.1\%$, while for $30$ projections it was $28.17\% - 40\%$. Other parameters such as {\it Tb.Th} was affected significantly as well for the two different numbers of projections. The {\it Tb.Th} decreased by up to $43.2\%$ of its baseline value (from $0.37$ mm to $0.21$ mm or from $0.34$ mm to $0.24$ mm).

The results from the CSDS algorithm show that the differences in the parameters are relatively smaller than those of the FDK method. For instance Tb.Sp increased by up to $14.6\%$  difference of its baseline value for $50$ projections and $5.71\% - 11.27\%$ for $30$ projections. This is due to the absence of noise speckles in the binary images shown in figure~\ref{fig:binary images}. It is reported as well that {\it Tb.Th} increased only up to $5.88\%$.

% Due to the appearance of the noise speckles induced by the noise on the binary images, Tb.Th and Tb.Sp values decreased dramatically for both samples as it was reported in Table~\ref{healthy parameter} and Table~\ref{OA parameter}. 

The {\it BV/TV} increased by a relatively small amount: from CSDS method it increased up to $6.06\%$, not significant difference compared to the FDK method for which the deviation was up to $5.66\%$.

Finally, reduction of the number of projections had less significant effects on the binary images of the CSDS method. The appearance of the speckles noise was insignificant even for a small number of projection images. Therefore, better results in the thresholded images were obtained. 
%There are almost no show of speckles on the binary images. 
As we can see in Table~\ref{healthy parameter} and Table~\ref{OA parameter}, the trabecular bone morphometric parameters calculation ({\it BV/TV}, {\it Tb.Th}, and {\it Tb.Sp}) for healthy and OA samples using sparser projection images is relatively closer to the values of the baseline parameters.

The results show that implementing the CSDS approach to reconstruct the inner structure of the samples using considerably sparse projection images outperforms the conventional FDK approach. The CSDS reconstructions seem to be smoothing out the edges, however by increasing the scale parameter in the shearlet transform, we should be able to capture more details of the image.

Despite its success, the computational burden of the CSDS method is relatively high. However, the computation time could be sped-up by implementing parallelized GPU code. Another acceleration strategy is to compute the shearlet decomposition in a serialized manner so that one does not need to keep all shearlet coefficients in memory at the same time. The computation times of the FDK and the shearlet based method using different number of projection images are shown in Tables~\ref{Shearlet Computation time} and \ref{FDK Computation time}. %The  shearlet-based method were run with stopping rule criterion and the number of iterations of each  computation is reported in table~\ref{FDK Computation time}.

In this study, there is no statistical comparisons because in fact collecting {\it in vitro} samples from patients is relatively hard and time consuming. Therefore, the results presented here were reported as a preliminary study. However, for the future work, statistical analysis in comparing the morphometric variables for each samples might be also considered. The range of {\it a priori} sparsity level $\mathcal{C}_{pr}$ from more data could also be computed. Applying the method {\it in vivo} would be also interesting to do as a future research as high ionizing radiation doses in $\mu$CT in {\it in vivo} experiment could be reduced. It has been discussed that high radiation could increase the risk of cancer, birth defects or heritable mutations \cite{brenner2003cancer,mitchel2007low}.

\section*{Acknowledgements}

This work was supported by the Academy of Finland (Finnish Centre of Excellence in Inverse Problems Research 2012--2017, and grant nos. 268378 and 303786), and the European Research Council under the European Union’s Seventh Framework Programme (FP/2007-2013)/ERC Grant Agreement no. 336267. We warmly thank Maximilian M{\"a}rz and Tatiana Bubba for discussions about the shearlets.

\bibliographystyle{IEEEtran}

\bibliography{IEEE_InverseProblems}

\end{document}